\documentclass[manuscript,nonacm]{acmart}
\usepackage{graphicx}

\AtBeginDocument{%
  }

\usepackage{amsmath}
\usepackage{xspace}
\usepackage{subcaption}

\usepackage{multirow} 
\usepackage[table]{xcolor}  
\usepackage{colortbl}       %

\usepackage{enumitem} 
\usepackage{tabularx,tabulary}

\usepackage{xspace}

\usepackage{siunitx} 

\usepackage{tcolorbox}
\usepackage{hyperref}
\usepackage{arydshln}

\MakeRobust{\ref}

\makeatletter
\newcommand{\labeltext}[2]{%
  \@bsphack
  \csname phantomsection\endcsname 
  \def\@currentlabel{#1}{\label{#2}}%
  \@esphack
}
\makeatother

\newtcbox{\highlight}[1][magenta]{on line, arc=0pt,colback=#1!10!white,colframe=#1!50!black, before upper={\rule[-3pt]{0pt}{10pt}},boxrule=1pt, boxsep=0pt,left=4pt,right=3pt,top=2pt,bottom=1pt}

\newcommand{\ccol}[1]{\cellcolor{blue!20}\textbf{#1}}

\newcommand{\tool}{\textit{Autonomy Design Assistant}\xspace}

\begin{document}

\title[Towards Informed Autonomy Design]{Towards Informed Autonomy Design: Understanding the Interplay of Function Characteristics and Autonomy Preferences in Automotive Agents}

\author{Sebastian Zepf}
\authornote{Both authors contributed equally to this research.}
\orcid{0000-0002-1268-146X}
\affiliation{%
  \institution{Mercedes-Benz AG}
  \city{Böblingen}
  \country{Germany}}
\email{sebastian.zepf@mercedes-benz.com}

\author{Annika Stampf}
\authornotemark[1]
\orcid{0000-0001-5539-4957}
\affiliation{%
  \institution{Ulm University}
  \city{Ulm}
  \country{Germany}}
\email{annika.stampf@uni-ulm.de}

\author{Markus Schuster}
\orcid{0009-0000-7856-8577}
\affiliation{%
  \institution{Ulm University}
  \city{Ulm}
  \country{Germany}}
\email{markus.schuster@uni-ulm.de}

\author{Enrico Rukzio}
\orcid{0000-0002-4213-2226}
\affiliation{%
  \institution{Ulm University}
  \city{Ulm}
  \country{Germany}}
\email{enrico.rukzio@uni-ulm.de}

\author{Mark Colley}
\orcid{0000-0001-5207-5029}
\affiliation{%
 \institution{University College London}
 \city{London}
 \country{UK}}
\email{m.colley@ucl.ac.uk}

\renewcommand{\shortauthors}{Zepf, Stampf et al.}

\begin{abstract}

  Proactive, autonomous interactions receive high interest in the field of intelligent assistants, including the automotive domain. A central challenge, however, is identifying appropriate levels of autonomy for different functions while maintaining user acceptance and trust. This work examines how function characteristics relate to suitable autonomy levels and whether these relationships generalize across functions. We conducted an expert workshop to define autonomy levels and identify relevant function characteristics. We then collected data from 126 participants, who rated 14 in-vehicle functions regarding autonomy preferences and function characteristics, from which we trained machine learning models to predict appropriate autonomy levels. Building on our findings, we developed our prototypical \tool that supports designers in making autonomy-related decisions, which we evaluated qualitatively with nine experts. Our results suggest that function characteristics can rank functions by their suitable level of autonomy and thereby guide more informed autonomy design decisions, whereas predictions for individual users remain uncertain.
\end{abstract}


\begin{teaserfigure}
  \centering
  \includegraphics[width=\textwidth]{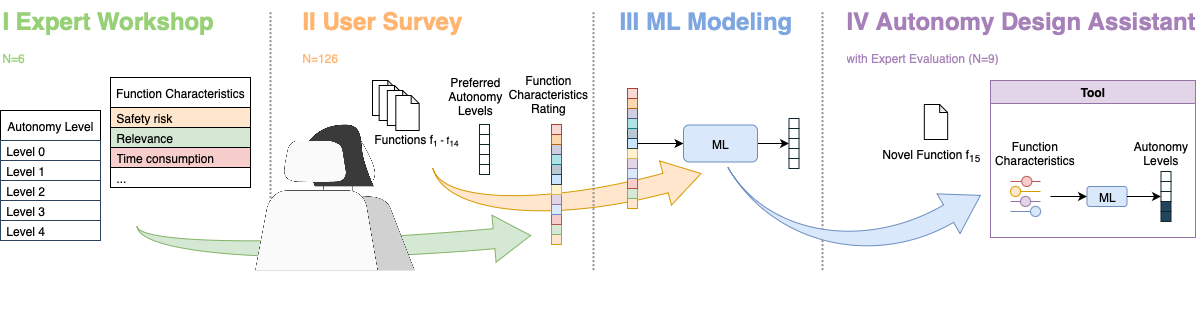}
  \caption{
Our multi-step approach to inform autonomy design for vehicle functions. 
(I) An expert workshop (N=6) defined five autonomy levels and twelve function dimensions. 
(II) An online survey (N=126) captured autonomy preferences and dimension ratings for 14 functions. 
(III) We trained multi-label classification models (e.g., CatBoost) to predict suitable autonomy levels based on function characteristics. 
(IV) The interactive \tool enables designers to rate function characteristics and explore model predictions; it was qualitatively evaluated with N=9 experts.
}
  \Description{The figure is a left-to-right visual summary of the research workflow, divided into four labeled stages:
I. Expert Workshop (green): A table shows five autonomy levels (Level 0–4) and a list of function characteristics such as safety risk and time consumption.
II. User Survey (orange): An abstract human figure faces a set of functions. For each function, participants rated autonomy preferences and function characteristics.
III. ML Modeling (blue): A machine learning (ML) box is shown receiving these ratings and producing predicted autonomy levels.
IV. Prototype Tool (purple): A graphical interface allows users to input characteristics of a novel function, which the ML model uses to predict appropriate autonomy levels. Arrows connect all four stages, showing the flow from expert input to survey data, machine learning, and final prototype usage.}
  \label{fig:teaser}
\end{teaserfigure}

\maketitle

\section{Introduction}
Intelligent assistants enable new functionalities and enhance user experiences, thus becoming key differentiators in the automotive industry~\cite{lee2022systematic}. Numerous car manufacturers developed on-board assistants, such as Mercedes' MBUX Virtual Assistant~\cite{mbux2022} and Renault's digital avatar Reno~\cite{reno}. These assistants, commonly known as conversational agents, are typically activated by user requests~\cite{meurisch2020exploringUEX}, to which they promptly respond~\cite{xi2023preprint}. These agents primarily perform simple, reactive tasks~\cite{meck2023may, meck2021design, du2024towards}. Advances in Artificial Intelligence (AI), particularly in Large Language Models, have significantly enhanced the reasoning and anticipation capabilities of AI models \cite{zhang2024llm}. This progress allows creating assistive systems that can initiate interactions and actions proactively, thereby improving efficiency and effectiveness. Current commercial applications in automotive~\cite{mbux2022} as well as in smart home applications~\cite{alexaplus2025} still mainly focus on proactive suggestions, while AI agents' function-calling capabilities can enable autonomous action execution based on context (e.g., user states~\cite{stampfImplicitInteractionHighly2022}) through task-specific functions or API calls, allowing AI systems to anticipate user goals and act on behalf of the user~\cite{berube2024proactive}.

Proactive and autonomous agents offer promising opportunities, such as increased task and interaction effectiveness and efficiency, which can enhance comfort and reduce cognitive load for non-driving related tasks~\cite{schmidt2020users}, enabling users to maintain focus on driving. Further, users desire proactivity in intelligent in-vehicle systems \cite{schmidt2018survey}. Consequently, proactive systems are extensively studied in the automotive domain~\cite{semmens2019now, wu2021learning, mathis2023proactive, schmidt2018survey}. However, significant concerns exist regarding appropriateness in specific contexts, loss of control, privacy~\cite{reicherts2021may, meurisch2020exploringUEX}, and interference in multi-party settings~\cite{zargham2022understanding}. As one aspect of appropriateness, several works addressed the question of \textit{when to interrupt} the user proactively \cite{semmens2019now, wu2021learning, mathis2023proactive}. Further, \citet{meck2023may} investigated \textit{how to communicate} when being proactive to meet user preferences. Another aspect is the question, \textit{which functions} should be proactive. Typically, research on proactivity in the automotive domain leverages exemplary use cases as a basis for investigation. However, existing work suggests that user preferences regarding proactivity highly depend on the use case \cite{schmidt2019exploration, mathis2023exploring}. Accordingly, the generalizability of use-case-specific findings remains unclear, limiting their scalability to other use cases. Further, the question of \textit{how much autonomy} users prefer for considered use cases remains open. 

Supported by initial indications that users prefer use cases that improve efficiency and comfort to be proactive~\cite{mathis2023exploring}, we hypothesize that the characteristics of a function, such as efficiency and safety-relevance, determine whether and to what extent users desire autonomy. We investigate the interplay between relevant function characteristics and the levels of autonomy (LoA) users consider suitable for automotive functions. The goal is to support designers in early design phases of proactive and autonomous systems with initial guidance on suitable autonomy levels for new use cases by assessing relevant function characteristics. We explore the following research questions (RQs):

\begin{itemize}[noitemsep, leftmargin=10pt]
    \item RQ1: Which LoA are relevant for in-vehicle functions and which function characteristics are pertinent for the appropriateness of LoA?
    \item RQ2: How well can we predict suitable LoA from function characteristics?
    \item RQ3: How to support designers in finding suitable LoA?
\end{itemize}

We conducted a workshop with N=6 experts in the automotive and intelligent user interface domain, an online survey (N=126), and designed, implemented, and evaluated (N=9 experts) the \tool (see \autoref{fig:teaser}). 

\noindent\highlight[purple]{Contribution Statement~\cite{wobbrockResearchContributionsHumancomputer2016}}
\begin{itemize}[noitemsep, leftmargin=11pt]
\item \textbf{Theoretical:} We identify and define key characteristics of in-vehicle functions, such as safety risk, increased safety, and privacy, influencing how much autonomy users prefer. We propose that these characteristics are a basis for generalizing autonomy preferences across different functions.

\item \textbf{Empirical:} We present findings from an expert workshop (N=6) and a user survey (N=126), collecting autonomy preferences for 14 in-vehicle functions and ratings of their characteristics. Using this data (1,764 samples), we train ML models to predict LoA based on function characteristics.

\item \textbf{Artifact:} The interactive \tool allows designers to explore predicted LoA for new functions based on function characteristics ratings. \tool was assessed via an expert evaluation (N=9).
\end{itemize}

\section{Related Work}
Autonomous systems can perceive their environment via sensors to act proactively and appropriately~\cite{du2024surveycontextawaremultiagentsystems}, aiming to execute decisions and actions without constant user input. Thus, autonomy is defined by the system’s ability to achieve a given goal independently, without human intervention~\cite{meurisch2020exploringUEX, xi2023preprint}. Modern approaches to system autonomy leverage ML to adapt to unforeseen and dynamic scenarios~\cite{endsley2017autonomy}. Proactive behavior is characterized by two fundamental properties: acting on behalf of the user and initiating tasks independently~\cite{salovaara2004proactivemanagement, kraus2021roleoftrust}, i.e., approaching the user without waiting for a command. Accordingly, proactivity adds a layer of initiative to autonomy, but requires some autonomy. 

In the automotive domain, proactive and autonomous agents promise to optimize driver support and vehicle interaction~\cite{semmens2019now, wu2021learning, mathis2023proactive, schmidt2020users}. \citet{schmidt2018survey} found in their online survey that users appreciate proactivity in intelligent systems. In a driving simulator study~\cite{schmidt2020users}, users generally accepted proactive interactions, and their quick responses suggested that interactions required minimal cognitive load. One critical aspect is the timing of proactive interactions. \citet{semmens2019now} found through a driving experiment that approaching drivers just before or during driving maneuvers (e.g., turn-taking) or in phases with uncertainty about directions can cause stress. They demonstrated that leveraging accessible automotive data can help select optimal moments for interaction. \citet{wu2021learning} modeled opportune moments for in-vehicle proactivity using sensor fusion. The results indicate that automotive, physiological, and visual data allow predicting driver availability. Finally, \citet{mathis2023proactive} found in a simulator study that most situations do not affect the appropriateness of proactive interactions during automated driving, except for extreme scenarios such as an approaching emergency vehicle. They identified travel time and driver state as relevant factors.

Effective proactive support requires not only suitable timing but also a precise understanding of the appropriate LoA for each function and context, which is sparsely addressed in existing research. Definitions of proactivity and autonomy levels can vary across different frameworks and domains; for example, \citet{kraus2021roleoftrust} categorize four levels (none, notification, suggestion, and intervention), while \citet{meurisch2020exploringUEX} consider three classes (reactive, proactive, autonomous) and \citet{du2024towards} define a set of five classes differentiated by the amount of autonomy and assumptions made. \citet{beer2014toward} defined ten levels specifically for robots, based on the model by \citet{sheridan1978LOAs}, which describes ten different levels of automation ranging from fully automated decisions and actions by the system to scenarios where all decisions and actions are managed solely by humans. Accordingly, there is no clear consensus on a clear set of relevant levels yet.

Despite the potential of autonomous and proactive systems to take over human tasks, they face significant challenges. To ensure the effectiveness and prevent misuse, such as overreliance, careful design and specific user training are required~\cite{parasuraman1997humans, lee2008review, lee2004trust}. Moreover, ML models perceived as black boxes lack transparency for users~\cite{chen2023gap}. \citet{cramer2010people} emphasizes that while users may not fully understand the inner workings of adaptive and autonomous systems, they need sufficient comprehension to evaluate the system’s competence and provide necessary feedback for effective adaptation. Additionally, \citet{shneiderman2021humancenteredai} criticizes the Levels of Automation model, arguing that increasing autonomy reduces human control. This critique highlights the importance of determining the appropriate LoA for each function and context, as it directly impacts the system's effectiveness and user acceptance. However, the current state of the art lacks a generalizable approach for making informed decisions about appropriate LoA for new use cases from early design phases on -- each new use case requires extensive individual investigation and user research. 

To address this gap, we evaluate the impact of function characteristics on the appropriateness of certain LoA and approach the prediction of suitable LoA from function characteristics. Further, we propose a tool that enables designers to explore autonomy in new use cases.

\section{Expert Workshop}

We conducted an expert workshop to (1) define relevant LoA for the context of human-vehicle interaction and (2) identify characterizing dimensions of potential autonomous functions that impact which level of autonomy is appropriate. 

\subsection{Workshop Design}

First, a presentation-based introduction about autonomy and the goals was given to the participants. After the introduction, the remaining questions were answered to ensure common ground and a clear understanding of the objectives ($\approx$ 10 min). 
Then, we obtained informed consent regarding participation, video recording of the session, and demographic information, including age, gender, current occupation, and education ($\approx$ 5 min). Subsequently, all participants performed a 10-min brainstorming regarding vehicle functions they imagined to have high potential for proactivity and autonomous execution. After brainstorming, every participant presented their ideas. This icebreaker created an open atmosphere among participants and fostered creative thinking. 
Then, the first main task was executed, namely the definition of relevant LoA. This task was divided into four stages: 1.) Brainstorming, 2.) Clustering, 3.) Prioritization, and 4.) Finalization. In the brainstorming phase, every expert was asked to ideate which differentiations of autonomy should be considered and reference existing concepts for autonomy and proactivity. In the clustering phase, existing concepts and ideas were clustered based on similarity in system autonomy and user involvement to create one set of clearly differentiable levels. In the prioritization phase, every expert voted for each level that they considered important. Lastly, in the finalization phase, the voting result served as the basis for deriving the final LoA set, completing the first main task in $\approx$ 50 min. The second main task was to identify the most important characteristic dimensions of functions and categorize them into different LoA. Analogous to the first task, the process was divided into 1.) Brainstorming, 2.) Clustering, 3.) Prioritization, and 4.) Finalization. The brainstorming used the functions gathered in the introduction task as a basis. This second main task took $\approx$ 55 min, resulting in an overall workshop duration of around 2h. 

The workshop was conducted online using Zoom to include experts from different locations. For collaboration, the visual collaboration platform \href{https://www.mural.co/}{Mural} was used. We involved six experts from academia (3) and industry (3) for diverse perspectives. Participants were M=30.17 years old (SD=4.02, range: 26 to 35), and the group involved four males and two females. Three participants had doctorates, and three had master’s degrees. Participants held backgrounds spanning interaction design, human factors research, and computer science, with the roles: UX designer, research associate, postdoctoral researcher, and HCI lecturer, bringing expertise from both academic and applied perspectives on in-vehicle systems. In their field, they had an experience of M=5.17 (SD=3.19) years. 

\subsection{Results}
To analyze the workshop results, we first transcribed the workshop conversation. Then, one author reviewed the full transcripts, while two additional authors independently reviewed the workshop board and post-it notes. All three authors created initial codes for LoA and FCDs. Through iterative discussions, the codes were refined and merged into a final, consistent set.
The workshop yielded two primary outcomes: 1) a diverse set of proposed LoA and 2) a set of FCD to characterize and evaluate functions across these levels.

Regarding the LoA, participants named existing conceptualizations from other fields such as human-computer interaction~\cite{meurisch2020exploringUEX}, mobility and transport studies~\cite{hopkinsTalkingAutomatedVehicles2021}, and science and technology studies~\cite{staytonItsTimeRethink2020}.
Their suggestions included conventional levels like driver support and full autonomy, as well as more nuanced forms such as cooperative, negotiable, or adaptive autonomy. They also highlighted the importance of perspective (e.g., technological vs. user-centric), the evolution of autonomy over time, and situation-specific system behavior.




Regarding the FCD, participants proposed and clustered aspects to characterize functions. Eight primary clusters emerged, namely \textit{risk and safety considerations}, \textit{relevance and urgency}, \textit{personal versus external perspectives}, \textit{personalisation and user adaptation}, \textit{system capability and reliability}, \textit{privacy and social considerations}, \textit{mental and physical demands}, and \textit{interaction and time efficiency}.
Additional aspects were raised during open discussion, such as the \textit{user’s state} (e.g., perception of the current situation), \textit{spatial context} (e.g., road environment and surrounding traffic), interaction \textit{"magicality"} (i.e., the execution of an action that exceeds the user’s expectations in a way that feels effortless and extraordinary, creating a sense of surprise), and the broader \textit{context of autonomy}. These contributions reflect the group's interdisciplinary perspectives and their shared focus on designing context-aware, user-sensitive autonomy.




\subsection{Final Autonomy Levels and Function Characteristic Dimensions}

In a series of follow-up discussions, three authors revisited the workshop results to finalize the proposed LoA and FCD. 

\subsubsection*{Levels of Autonomy.}

An analysis of existing models and participant ideas led to a refined model with five levels:
\begin{itemize}[left=0pt,labelsep=0.25em]
    \item \textbf{Level 0:} User performs task without system support (reactive)
    \item \textbf{Level 1:} User performs task with system support (preselection/ suggestion)
    \item \textbf{Level 2:} System performs task with user approval
    \item \textbf{Level 3:} System performs task pending user veto
    \item \textbf{Level 4:} System performs task
\end{itemize}

\subsubsection*{Function Characteristic Dimensions}

Twelve FCD were identified based on the workshop findings: 

\begin{itemize}[left=0pt,labelsep=0.25em, noitemsep]
       \item \textbf{Safety Risk:} The execution of the function can lead to a safety-relevant risk.
    
    \item \textbf{Increased Safety:} The execution of the function may improve safety for the user or vehicle occupants.
    
    \item \textbf{Relevance:} The execution of the function is meaningful and useful in the current situation from the user’s perspective.
    
    \item \textbf{Magicality of Interaction:} Action execution can feel magical, surprising the user, and exceeding their expectations.
    
    \item \textbf{Privacy:} The function execution may involve a privacy violation (e.g., access to/disclosure of personal information).
    
    \item \textbf{Trust:} The user needs to have trust in the vehicle/system to accept the execution of the function.
    
    \item \textbf{Time Consumption:} The function can be executed quickly or slowly from the user’s perspective.
    
    \item \textbf{Repetitiveness:} The function is typically performed repeatedly or habitually in everyday situations.
    
    \item \textbf{Situational Context:} The function relevance depends on the current situation (e.g., location, system state, user state).
    
    \item \textbf{Social Risk:} Function execution may result in social consequences such as embarrassment or social disapproval.
    
    \item \textbf{Urgency:} The function needs to be performed immediately to avoid negative consequences or to achieve a benefit.
    
    \item \textbf{Complexity:} The function requires high cognitive and/or physical effort to execute.
\end{itemize}




\section{Online Survey}

As the second step towards a generalized framework for autonomy level categorization, we conducted an online survey to gather quantitative data from users regarding their preferred LoA for exemplary functions and their subjectively perceived importance of the proposed FCD. The objective was to examine whether there is a correlation between the characteristics of the function and the resulting user preferences regarding proactivity that allows a generalized inference of suitable proactivity from function characteristics. 

\subsection{Survey Design}

To investigate the generalizability of autonomy decisions based on function characteristics, an online survey via LimeSurvey was conducted to reach a broader audience. 

We defined an exemplary set of common vehicle functions chosen to represent a broad range of different function types, such as comfort functions, entertainment functions, communicative functions, informative functions, and driving-related functions. The selection of functions was built upon these categories, the findings from our expert workshop, and existing work on proactivity in the automotive context~\cite{du2024towards}, resulting in the following set of functions: Adjust seat positioning, Adjust in-vehicle temperature, Send a text message, Start a phone call, Navigation control, Change driving mode, Select parking space, Overtake vehicle ahead, Start driving, Provide weather update, Adapt ambient light, Change song, Provide traffic news, and Start a movie.
The survey was composed of an introduction and four survey sections. 

After giving consent, the first survey section requested users to report their preferred LoA for each of the fourteen selected functions. As choices for each function, the users were provided with the five LoA defined in our expert workshop. While participants were intentionally confronted with the functions without giving contextual scenarios (i.e., using the method of random situation design~\cite{rauthmann2015principles}), participants could select one or several LoA for each function; most multiple selections formed a contiguous range (30 of 1,764 selections were non-contiguous). 
In the second part of the main survey, participants rated the twelve FCD for each of the fourteen functions. For each FCD, participants were provided a one-sentence statement and a 5-point Likert scale to report how much they agree with the statement considering the given function. The Likert scale comprised the levels strongly disagree, disagree, neutral, agree, and strongly agree.  
Finally, the fourth section combined demographics questions (gender, age, current occupation) and questionnaires capturing information about the users that might relate to their ratings, namely the Propensity to Trust subscale of the Trust in Automation questionnaire~\cite{korber_theoretical_2018} and the short form of the Big Five Inventory~\cite{soto2017short}. Overall, the survey procedure took $\approx$40 min and included five attention checks. 



Participants were recruited through the trial administration portal of our institution, the networks of the contributors of this research, Reddit forums, and Prolific. 
Participants had to be 18 years old and have a valid driver's license.

\subsection{Results}

The survey lasted $\approx$5.5 weeks, with 126 valid responses. 
Participants were M=30.66 years old (SD=10.15, range: 18 to 66), with 65 men, 55 women, four non-binary individuals, and two participants not providing information about their gender. Most participants were employed (n=64), followed by students (n=48), self-employed individuals (n=8), one pupil, and one unemployed participant. Four participants answered "other." Most participants resided in Germany (n=115), followed by Austria (n=6) and Switzerland (n=5). They held a driver's license for an average of 11.94 years (SD=10.06, range: 1 to 48). Regarding driving frequency, 31 participants reported driving daily, 14 for 5-6 times a week, 23 for 3-4 times a week, 20 for 1-2 times per week, 26 for 1-3 times per month, and 12 for less than once per month. 



\begin{figure}
    \centering
    \includegraphics[width=0.75\linewidth]{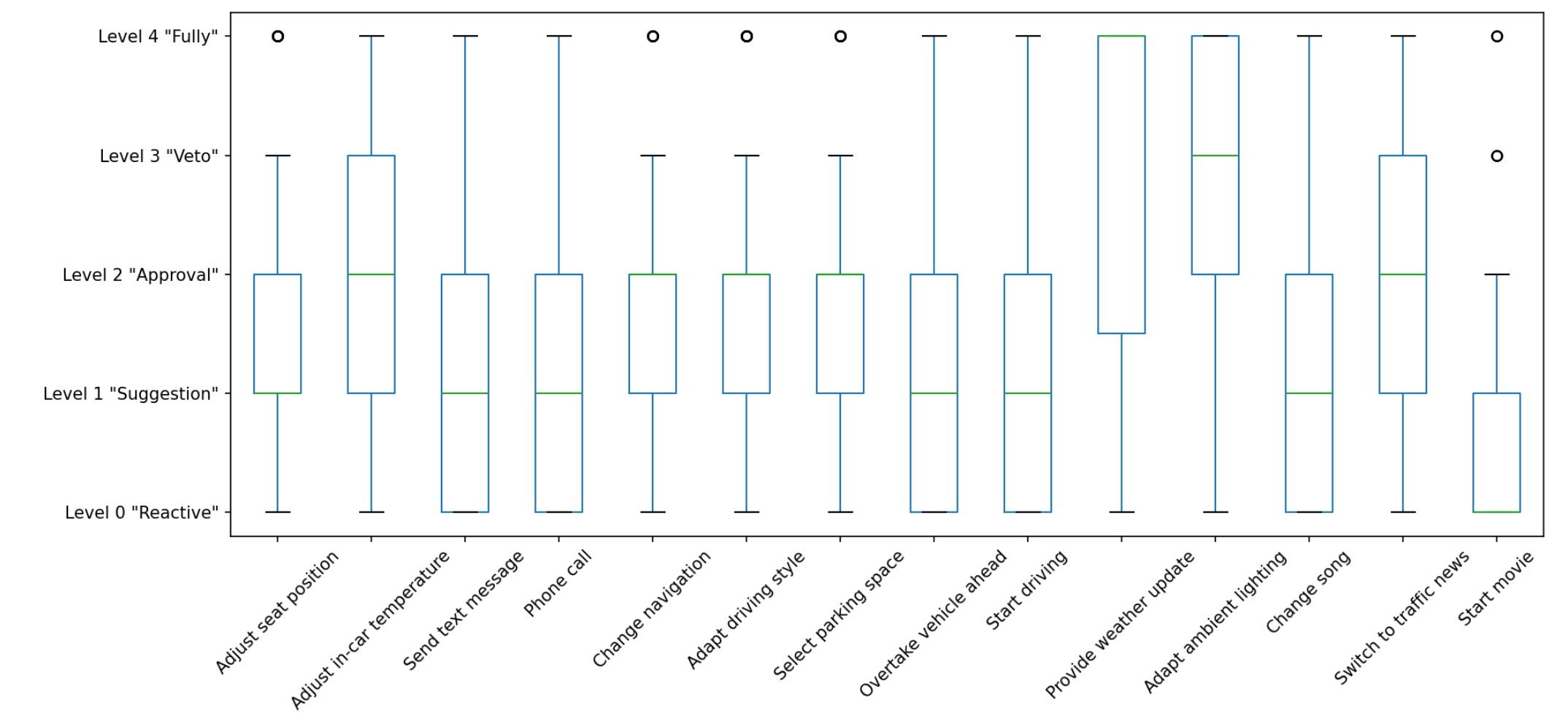}
    \caption{User autonomy level ratings per function.}
    \label{fig:autonomy_actions}
    \Description{Box-plots per considered function showing the mean and standard deviation of the Level of Autonomy ratings provided by the survey participants. Provide a weather update. Adapt ambient lighting received the highest average autonomy rating, while Start movie ranks last.}
\end{figure}

\subsection{Levels of Autonomy (LoA)}

There was a general preference for lower LoA (see \autoref{fig:autonomy_actions}), with most actions averaging between Level 1 and Level 2. Participants favored higher control, either with system suggestions or approval required for execution. Notable exceptions included higher average scores for actions like "Provide weather update" (2.77) and "Adapt ambient lighting" (2.48), while "Start movie" received a notably low average rating of 0.53.

Providing a weather update (2.77) and adapting the ambient lighting (2.48) received the highest ratings, followed by providing traffic news (1.77) and driving-related and comfort functions such as changing the driving mode (1.73), navigation control (1.72), selecting a parking space (1.71), adjusting the temperature (1.66), and the seat position (1.58). ``Overtake vehicle ahead'' ranked in the middle (1.48). Communication and entertainment functions received the lowest ratings (starting a phone call 1.25, sending a text message 1.15, changing the song 1.11, starting a movie 0.53), as did ``Start driving'' (1.07).
The minimum and maximum values for each action showed substantial variation in preferences, with boxplots indicating no single level was distinctly favored. Standard deviations ranged from 0.92 to 1.43, reflecting acceptable variance in preferences.

\subsection{Function Characteristic Dimensions}

\begin{figure}
    \centering
    \includegraphics[width=0.9\linewidth]{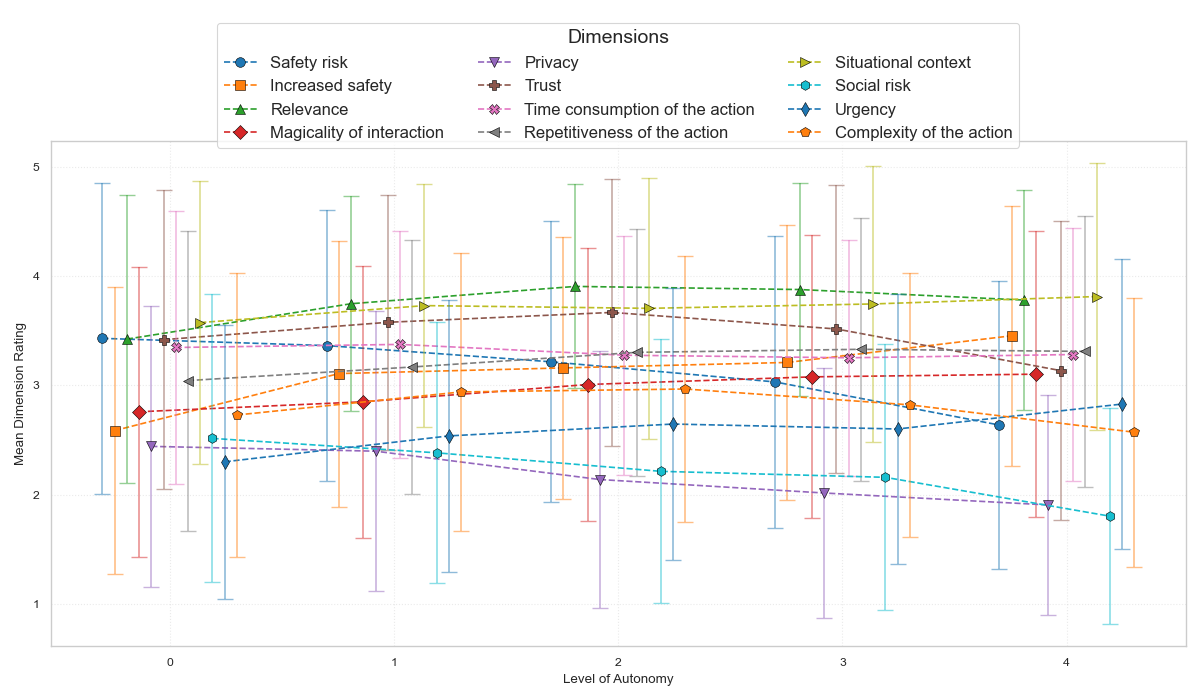}
    \caption{Mean rating per FCD across autonomy levels, including standard deviation.}
    \label{fig:trendline_dimensions}
    \Description{It shows the mean rating per function characteristic dimension (FCD) for each autonomy level, including standard deviation. Three of the twelve FCD (Safety risk, Social risk, and Privacy) show strictly decreasing mean ratings with increasing LoA. Two FCD show strictly increasing mean ratings with increasing LoA, in particular Increased safety and Magicality of the interaction. The other FCDs do not show a clear trend.}
\end{figure}

\autoref{fig:trendline_dimensions} illustrates the mean rating for each FCD across LoAs. 3/12 FCD ("Safety risk," "Social risk," and "Privacy") show strictly decreasing mean ratings with increasing LoA, indicating that functions with potential physical or social harm for the user are connected to lower autonomy preferences. Two FCD show strictly increasing mean ratings with increasing LoA: "Increased safety" and "Magicality of the interaction." For "Increased safety," this observation aligns well with the potential of function execution to improve safety and thus serves as a counterpart to "Safety risk." 
For the FCD of "Urgency" and "Repetitiveness of the action," we observe a visual tendency for higher mean ratings with increasing LoA. However, these trends are not strict, involving a decrease from level two to level three for "Urgency" and a stagnation from level two to level four for "Repetitiveness of the action." Finally, there are five FCD without a clear trend. "Trust" and "Complexity of the action" show the highest mean ratings for medium autonomy (levels one to three), the mean ratings for relevance increase up to level two and slightly decrease with levels three and four, and both "Time consumption of the action" and "Situational context" involve multiple changes between increase and decrease.

\section{Generalized Mapping of Dimensional Ratings to Autonomy Levels}

Based on the survey data, we examined the potential of predicting the appropriate LoA for a given function based on the FCD evaluation. As an approach, we chose classification models, using the LoA as target classes and the FCD as input features. We further chose a multilabel approach~\cite{tsoumakas2008multi} to allow output ranges or several output classes instead of only a single classification. We considered five different classification models, namely Random Forest, Cat-Boost, XGBoost, Support Vector Machine (SVM), and TabPFN~\cite{hollmann2025tabpfn}. 

\subsection{Experiments}
Random Forest~\cite{breiman2001random}, CatBoost and XGBoost~\cite{bentejac2021comparative}, SVM~\cite{mammone2009support}, and TabPFN~\cite{hollmann2025tabpfn} are popular classification models relatively robust to overfitting. 
The dataset consisted of 126 $\times$ 14 rows, totaling 1,764 samples. For the LoA, values were converted to binary: 1 if the autonomy level was selected and 0 if not. For the FCD ratings, the following values were applied: 1 = Strongly Disagree to 5 = Strongly Agree. 

\begin{table}[ht]
\footnotesize
\centering
\begin{tabular}{lccccc|lccccc}
\toprule
Level & 0 & 1 & 2 & 3 & 4 & Levels selected & One & Two & Three & Four & Five \\
\midrule
Frequency & 575 & 465 & 534 & 258 & 245 & Frequency & 1562 & 127 & 48 & 18 & 9 \\
\bottomrule
\end{tabular}
\caption{Frequency of levels and ranges in the dataset}
\label{tab:level-range-frequency}
\end{table}

We randomly split the samples into a training set (70\,\%, 1,234 samples) and a held-out test set (30\,\%, 530 samples). Each model was trained as a set of binary classifiers, one per autonomy level (binary relevance; scikit-learn's \texttt{MultiOutputClassifier}, and \texttt{OneVsRestClassifier} for the SVM). Hyperparameters were selected by an exhaustive grid search with five-fold cross-validation on the training set, using the macro-averaged F$_1$-score as criterion; TabPFN was used without tuning, as intended by its design. All classifiers predict a level if its predicted probability exceeds 0.5. We report all results on the test set (scikit-learn 1.9.1, CatBoost 1.2.10, XGBoost 3.4.1, TabPFN 9.2.0 with the TabPFN-v2 weights; grids and scripts in our repository).


To examine which FCD drive the predictions, we computed the permutation importance of each feature on the test set (30 repetitions, macro-F$_1$).

\subsection{Results}
\autoref{tab:classification_performance} summarizes the results of our experiments. 
CatBoost (M-Avg F$_1$=41\,\%, W-Avg F$_1$=45\,\%) and XGBoost (M-Avg F$_1$=42\,\%, W-Avg F$_1$=45\,\%) yield the best prediction performance and do not differ significantly (difference in macro-F$_1$: $-0.9$ percentage points, 95\,\% CI [$-2.5$, $0.7$]). Random Forest (M-Avg F$_1$=40\,\%, W-Avg F$_1$=43\,\%) and SVM (M-Avg F$_1$=39\,\%, W-Avg F$_1$=42\,\%) perform slightly lower. TabPFN reaches only M-Avg F$_1$=9\,\% (W-Avg F$_1$=12\,\%) because it was evaluated at the default decision threshold of 0.5 without class weighting and therefore rarely predicts a level (macro recall 6\,\%). Its ranking quality is comparable to the other models (macro ROC-AUC 0.63), and with per-level decision thresholds tuned for F$_1$ by five-fold cross-validation on the training set, TabPFN reaches M-Avg F$_1$=42\,\% (W-Avg F$_1$=45\,\%). All models except TabPFN show their lowest F$_1$-scores for Levels 3 and 4, the two least frequent levels (14\,\% of the samples each). This mainly reflects their low prevalence: in terms of ROC-AUC, Level 4 is the best-separated level for all models (0.71--0.74), whereas Level 3 is the hardest to separate (0.53--0.56).
While the overall classification performance with F$_1$-scores around 40\,\% is modest, it has to be compared with trivial strategies: predicting every autonomy level for every sample yields a macro-averaged F$_1$ of 37.4\,\% and a weighted F$_1$ of 41.5\,\% on the same test set (random guessing according to the level frequencies: 23.5\,\% and 26.3\,\%). CatBoost and XGBoost exceed the every-level baseline by 3.8 (95\,\% CI [1.6, 6.1]) and 4.7 ([2.7, 6.8]) percentage points macro-F$_1$ (participant-clustered bootstrap, 2,000 resamples); for Random Forest and SVM, the difference is not significant. The threshold-free macro ROC-AUC of all models lies between 0.62 and 0.64 (chance: 0.5), indicating weak but consistent patterns in the data.

\begin{figure}[ht]
    \centering
    \includegraphics[width=0.66\linewidth]{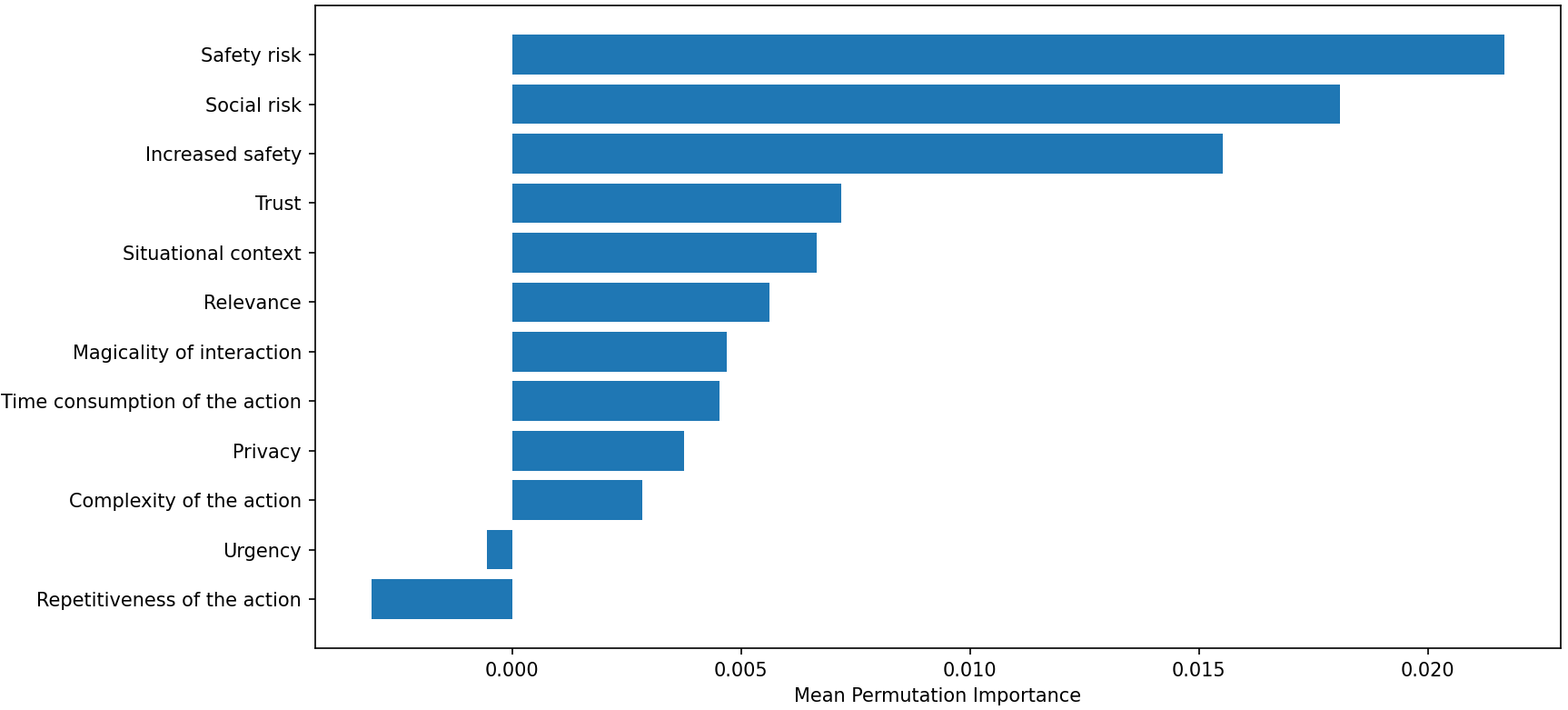}
    \caption{Mean permutation importance ($\pm$ SD over 30 repetitions, test set, macro-F$_1$) of each function characteristic dimension for CatBoost.}
    \label{fig:feature_importance}
    \Description{Mean permutation importance of each function dimension characteristic for the best classifier, CatBoost. The highest importance was Safety risk, followed by Social risk and Increased safety. There were two negative importances: Urgency and the even more negative repetitiveness of the action.}
\end{figure}

\autoref{fig:feature_importance} shows the feature selection results for the best-performing model, CatBoost. "Safety risk," "Social risk," and "Increased safety" are the most important features (mean permutation importance $>0.015$). \autoref{fig:feature_importance} shows the permutation importance of each FCD for CatBoost, computed on the test set (30 repetitions, macro-F$_1$). ``Safety risk'' (0.021), ``Social risk'' (0.015), and ``Increased safety'' (0.015) are the most important features. This aligns well with the previous observation that the mean ratings of these FCD change monotonically with the LoA (\autoref{fig:trendline_dimensions}; decreasing for safety and social risk, increasing for increased safety). The importances of ``Urgency'' and ``Repetitiveness'' are indistinguishable from zero (mean below one standard deviation), i.e., these FCD do not measurably contribute to the classification.

Because each level is predicted independently, the models predict on average 2.0 (Random Forest) to 3.1 (XGBoost) levels per sample, compared to 1.18 levels selected by participants, and 16--22\,\% of their predictions are non-contiguous (e.g., Levels 0, 1, and 4).

\subsection{Generalization to Unseen Functions}
In the random split, every participant and every function appears in both training and test data. A model that only knows which of the 14 functions a sample belongs to, without any FCD rating, already reaches a macro-F$_1$ of 42.6\,\% (weighted 44.6\,\%); the random split thus largely measures how well known functions are recognized. To evaluate the use case of the \tool{}, i.e., predicting autonomy for a function the model has not seen, we additionally applied leave-one-function-out cross-validation: each function was predicted by a model trained on the remaining 13 functions, using the hyperparameters selected above (\autoref{tab:generalisation}). On unseen functions, all models perform close to the every-level baseline (macro-F$_1$ 39.1--39.7\,\% vs.\ 37.5\,\%; ROC-AUC 0.60--0.62). At the level of functions, however, the predicted mean LoA correlates with the observed mean LoA across the 14 held-out functions (Spearman $\rho$ = 0.61--0.73, $p \le .02$), although the predictions are compressed towards medium levels (CatBoost: predicted means 1.42--2.03 vs.\ observed 0.51--2.81). A variance decomposition of the preferred LoA (mean selected level per participant and function) explains this pattern: 20\,\% of the variance lies between functions, 20\,\% between participants, and 60\,\% in participant-specific preferences for individual functions and noise.

\begin{table*}[t]
  \centering\footnotesize
  \begin{tabular}{lccc|cccc}
    \toprule
    & \multicolumn{3}{c|}{Random 70/30 split (as in \autoref{tab:classification_performance})} & \multicolumn{4}{c}{Unseen functions (leave-one-function-out)} \\
    Model & M-F1 & $\Delta$ M-F1 vs.\ every level [95\,\% CI] & AUC & M-F1 & $\Delta$ M-F1 vs.\ every level [95\,\% CI] & AUC & $\rho$ \\
    \midrule
    Every level (trivial) & 0.374 & -- & 0.50 & 0.375 & -- & 0.50 & -- \\
    Function identity only & 0.426 & -- & 0.66 & \multicolumn{4}{c}{not applicable (function unseen)} \\
    \midrule
    CatBoost & 0.412 & +0.038 [+0.016, +0.061] & 0.64 & 0.391 & +0.016 [+0.001, +0.031] & 0.61 & 0.61 \\
    XGBoost & 0.421 & +0.047 [+0.027, +0.068] & 0.63 & 0.397 & +0.022 [+0.012, +0.032] & 0.61 & 0.61 \\
    Random Forest & 0.400 & +0.026 [-0.003, +0.055] & 0.63 & 0.392 & +0.017 [+0.001, +0.034] & 0.62 & 0.63 \\
    SVM & 0.391 & +0.017 [-0.009, +0.044] & 0.62 & 0.391 & +0.016 [+0.000, +0.033] & 0.61 & 0.73 \\
    TabPFN$^\dagger$ & 0.415 & -- & 0.63 & -- & -- & 0.60 & 0.64 \\
    \bottomrule
  \end{tabular}
  \caption{Prediction performance relative to trivial baselines. Left: the random split of \autoref{tab:classification_performance},
  in which every participant and every function appears in both training and test data. Right: each of the 14 functions
  is predicted by a model trained on the other 13 (the use case of the \tool), pooled over all 1,764 samples.
  CIs: participant-clustered bootstrap (2,000 resamples). $\rho$: Spearman correlation between predicted and observed
  mean autonomy level across the 14 held-out functions (all $p \le .02$). $^\dagger$F1 with per-level tuned thresholds (test split).}
  \label{tab:generalisation}
\end{table*}


\begin{table*}[t]
  \centering\footnotesize
  \setlength{\tabcolsep}{5pt}
  \begin{tabular}{lrrr:rrr:rrr:rrr:rrr:r}
    \toprule
    \multicolumn{1}{c}{} & \multicolumn{3}{c}{Random Forest} & \multicolumn{3}{c}{CatBoost} & \multicolumn{3}{c}{SVM} & \multicolumn{3}{c}{TabPFN} & \multicolumn{3}{c}{XGBoost} & \multicolumn{1}{c}{Sup.} \\
    \cmidrule(lr){2-4}\cmidrule(lr){5-7}\cmidrule(lr){8-10}\cmidrule(lr){11-13}\cmidrule(lr){14-16}
    & \ccol{P} & \ccol{R} & \ccol{F1} & \ccol{P} & \ccol{R} & \ccol{F1} & \ccol{P} & \ccol{R} & \ccol{F1} & \ccol{P} & \ccol{R} & \ccol{F1} & \ccol{P} & \ccol{R} & \ccol{F1} \\
    Level 0 & 0.52 & 0.48 & 0.50 & 0.40 & 0.91 & 0.55 & 0.50 & 0.53 & 0.51 & 0.63 & 0.21 & 0.32 & 0.39 & 0.90 & 0.55 & 187 \\
    Level 1 & 0.37 & 0.67 & 0.48 & 0.30 & 0.94 & 0.45 & 0.31 & 0.67 & 0.42 & 0.00 & 0.00 & 0.00 & 0.31 & 0.93 & 0.47 & 144 \\
    Level 2 & 0.31 & 0.60 & 0.41 & 0.29 & 0.87 & 0.44 & 0.32 & 0.65 & 0.43 & 0.40 & 0.01 & 0.03 & 0.30 & 0.85 & 0.44 & 149 \\
    Level 3 & 0.18 & 0.47 & 0.26 & 0.16 & 0.38 & 0.23 & 0.14 & 0.43 & 0.21 & 0.00 & 0.00 & 0.00 & 0.17 & 0.54 & 0.26 & 72 \\
    Level 4 & 0.26 & 0.55 & 0.36 & 0.32 & 0.49 & 0.39 & 0.27 & 0.63 & 0.37 & 1.00 & 0.07 & 0.13 & 0.28 & 0.60 & 0.38 & 73 \\
    \hdashline
    M-Avg. & 0.33 & 0.55 & 0.40 & 0.29 & 0.72 & 0.41 & 0.31 & 0.58 & 0.39 & 0.41 & 0.06 & 0.09 & 0.29 & 0.76 & \textbf{0.42} & 625 \\
    W-Avg. & 0.37 & 0.56 & 0.43 & 0.31 & 0.80 & 0.45 & 0.35 & 0.59 & 0.42 & 0.40 & 0.08 & 0.12 & 0.31 & 0.82 & \textbf{0.45} & 625 \\
    \hdashline
    ROC-AUC (macro) & \multicolumn{3}{c}{0.63} & \multicolumn{3}{c}{0.64} & \multicolumn{3}{c}{0.62} & \multicolumn{3}{c}{0.63} & \multicolumn{3}{c}{0.63} & \\
    \bottomrule
  \end{tabular}
  \caption{Classification performance on the held-out 30\,\% (530 samples) of Random Forest, CatBoost, SVM, TabPFN, and
  XGBoost (binary relevance; hyperparameters selected by 5-fold grid search on the training 70\,\% using macro-F1):
  precision (P), recall (R), and F1 per autonomy level, macro- (M-Avg) and support-weighted (W-Avg) averages, and the
  threshold-free macro ROC-AUC. Support counts label occurrences. Predicting every level for every sample yields
  M-Avg F1 = 0.37 and W-Avg F1 = 0.41. TabPFN is evaluated at the default threshold of 0.5;
  with per-level F1-tuned thresholds it reaches M-Avg F1 = 0.42.}
  \label{tab:classification_performance}
\end{table*}

\section{\tool: An Interactive Tool for Exploring Autonomy of Functions}

To enable practitioners to explore suitable autonomy ranges for functions, including the impact of FCD, we designed an interactive tool called \tool that allows users to rate the different FCD and assess the LoA proposed by the best-performing classifiers, \textit{CatBoost} and \textit{XGBoost}.

\subsection{Concept and Implementation}

\begin{figure}[ht!]
    \centering
    \begin{subfigure}[t]{0.48\linewidth}
        \centering
        \includegraphics[width=\linewidth]{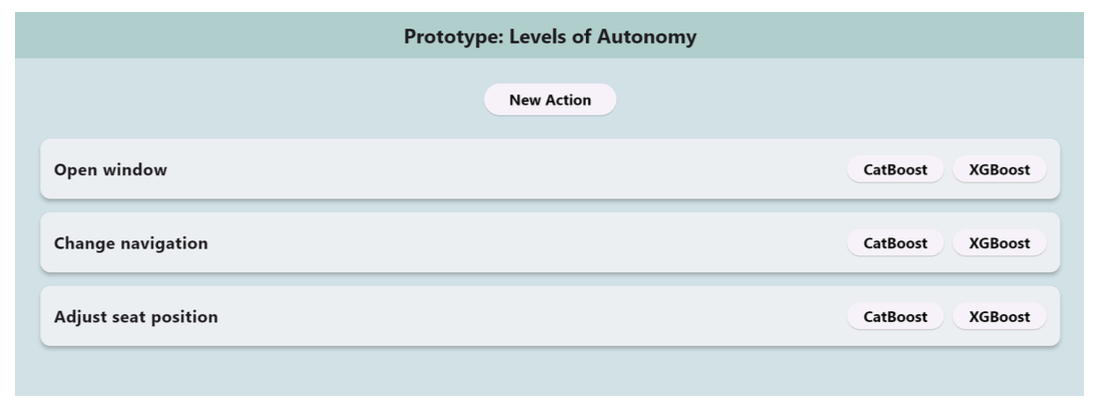}
        \caption{Start screen of \tool showing the already considered functions.}
        \label{fig:app_startscreen}
        \Description{This figure shows the prototypical implementation of the start screen. At the top, a "New Action" button is shown. Three actions are already present: "Open window", "Change navigation", and "Adjust seat position".}
    \end{subfigure}
    \hfill
    \begin{subfigure}[t]{0.5\linewidth}
        \centering
        \includegraphics[width=0.98\linewidth]{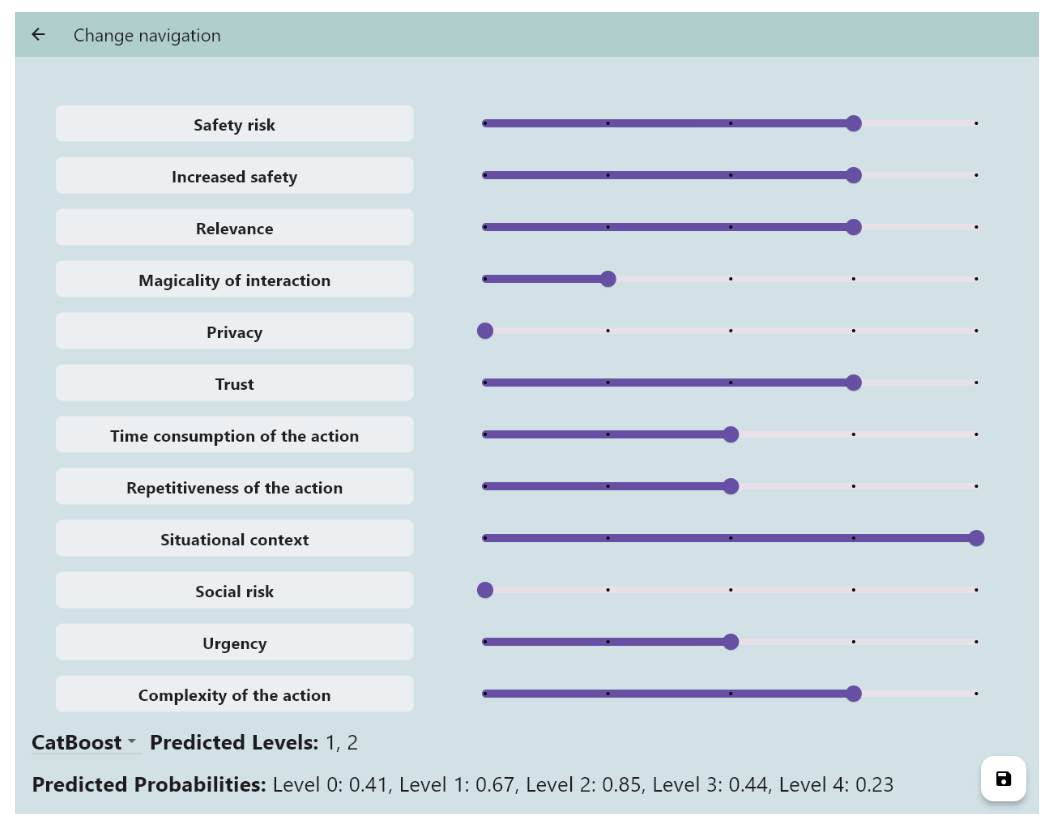}
        \caption{\tool's exploration screen showing available FCDs, dimension rating scales, and autonomy level prediction.}
        \label{fig:app_explorationscreen}
        \Description{For each of the FCDs, the user can adjust the values using sliders. At the bottom, the predicted LoA is shown including probabilities for all LoAs.}
    \end{subfigure}
    \caption{Prototype screens of \tool.}
    \label{fig:app_screens}
\end{figure}

The Flutter-based, thus platform-independent, \tool enables model application and performance testing on new data instances. It interactively creates new actions (functions), rates twelve FCD, and predicts the appropriate autonomy level based on these ratings. 
\autoref{fig:app_startscreen} and \autoref{fig:app_explorationscreen} show screenshots of the user interface.

Upon selecting an action classifier (the best-performing model, CatBoost or XGBoost), FCDs are listed on the left side, with tooltips displaying the corresponding definition statements as used in the survey. Sliders allow users to rate their agreement. 
Further, the predicted LoAs are shown, including probabilities for each level. Predictions update dynamically with slider adjustments or model selection.



\subsection{Expert Interviews}

To obtain insights into the potential impact on practitioners and their design process, we conducted expert interviews and a SWOT analysis to assess \tool. 


\subsubsection{Interview Procedure}

N=9 experts evaluated the \tool in a structured feedback process. We provided participants with standalone executable files and a detailed explanation. 
The experts added three example actions and rated the FCD for them. The predefined actions were: "Change mirror position,"
"Open window," and "Provide vehicle information." By assigning ratings to the FCDs of these three actions, experts could observe how individual FCDs influence the model’s predictions and compare the outputs of the two integrated models. Experts were then asked to complete a SWOT matrix. 

The experts had a mean age of 28.9 years ($SD = 4.17$, range: 24 to 35 years) and an average field experience with virtual assistants of 4.94 years ($SD = 2.88$, range: 1 to 10 years), including five female and four male participants. Five participants worked in academia, one in industry, and three in both.

\subsubsection{User Feedback}

\textbf{Strengths:} During the evaluation process, participants identified numerous strengths of the prototype. They highlighted that the tool is "easy to use" and enables users to quickly get a first estimate or pre-selection of autonomy levels as a basis for final design decisions. Especially the structured nature of the approach was appreciated, as was its ability to demonstrate the impact of different FCDs on LoA selection, thereby enhancing explainability. It was further stated that the tool enables more objective decisions by ensuring consideration of all influencing aspects (dimensions) and by providing the designer with data-driven insights. Using probabilities was regarded as beneficial, "estimating the uncertainty of the predictions" and facilitating expeditious, informed decision-making, while concurrently prompting critical reflection on the underlying probabilities. The approach was also valued for its ability to quantify complex decisions that are otherwise difficult to articulate. The FCD were found beneficial in evaluating options, with one participant specifically noting the thoughtful distinction between "Safety risk" and "Increased safety." Overall, the concept was described as a "great starting point" with "highly important and relevant" FCD and was regarded as a promising approach to set guidelines for a coherent user experience.

\textbf{Weaknesses:} 
Several weaknesses were also identified. One issue concerned the predicted LoA, which at times appeared inconsistent or unintuitive, as exemplified by the combination of Levels 0, 1, and 4. One participant regarded Level 1 as "too broadly defined." Additionally, the FCD scales
were not clearly visible, and a larger Likert scale (e.g., 7- or 10-point) was proposed to capture more nuanced responses. Participants further noted "that the dimensions are somewhat vague," which could lead to variance in the ratings between different experts. The most notable examples were "Social risk," "Privacy," and "Situational context." The latter was particularly critiqued for its excessive breadth and lack of specificity, given the considerable variation in the desired LoA that is known to occur depending on various contextual situations, such as the setting, urgency, or user state. Additionally, the FCDs were "rather task-focused and not interaction-focused," which may have resulted in the oversight of significant aspects of user experience. Furthermore, the interrelation of factors such as "Urgency" and "Safety risk" rendered them challenging to evaluate independently. Finally, the absence of an explanation for the model’s predictions and the limitations in providing guidance on how to interpret the FCD were seen as weaknesses. Participants proposed that clearer explanations could enhance understanding and more effectively "convey the intended message."

\textbf{Opportunities: } A range of opportunities for improvement and future development was also uncovered. Participants noted that \tool could be significantly more useful with a clearly defined and "more concrete set of dimensions." The incorporation of FCDs that reflect users’ desired level of control and technical affinity was suggested (referring to more differentiation between task-focused and interaction-focused autonomy). Furthermore, the presentation and explanation of the prototype were identified as a potential area for enhancement, with the suggestion to provide "examples for the individual levels," offer more detailed explanations of FCD, and enhance the reasoning behind predictions. The participants expressed a desire for enhanced explainability, particularly with regard to the rationale behind recommending a level. Additionally, using "a larger dataset" was regarded as a significant step towards enhancing \tool's reliability. Finally, it was suggested to include thought-provoking questions for each dimension and to offer a textual summary of the predictions, including arguments for or against specific levels of autonomy, as additional output.

\textbf{Threats: }
The experts also identified several threats. The subjective nature of the FCD ratings and potentially involved biases pose a considerable challenge. Participants noted that a "wrong weighting of the dimensions can lead to non-representative results." In addition, the currently used models were described as somewhat imprecise or vague, indicating a need for refinement to improve prediction clarity and accuracy. Utilising a "generally trained system might not perform well regarding individual set parameters," especially if FCD definitions are too broad and open to different interpretations. These points lead to the key concern of discrepancy between the expected and predicted LoA, underscoring the necessity for proper validation procedures. Additionally, the immediate presentation of predicted results during the FCD rating has been found to potentially prompt a process of "reverse-engineering" among users. 

\section{Discussion}
This work examined whether FCD can inform the assignment of LoA to in-vehicle functions and how well such a mapping generalizes to new functions. By combining qualitative expert input, quantitative survey data, and ML-based classification, we presented a multi-method approach to support practitioners in their autonomy design decisions. 

\subsection{Broader Context of Autonomy and Functions}
There is a wide range of autonomy and proactivity level models across the literature, yet no clear consensus on terminology or structure \cite{kraus2021roleoftrust, meurisch2020exploringUEX, du2024towards, beer2014toward}. \citet{du2024towards} describe a functionally similar model but refer to the levels as “proactivity levels.” Their approach, grounded in occupational and organizational psychology, distinguishes between autonomy and assumption. Assumption is the system’s capacity to anticipate user intentions. This overlap with our considered LoA highlights a need for greater alignment of autonomy level models across domains and suggests potential for interdisciplinary convergence.

Our participants preferred a lower LoA. Most frequently, Level 0 (user performs task without system support; 575 selections) and Level 2 (system performs task with user approval; 534) were selected, followed by Level 1 (465; see \autoref{tab:level-range-frequency}). Lower LoA were particularly preferred for functions that are unfamiliar or not typically used during manual driving (e.g., watching a movie). These preferences may also indicate a desire to remain in control. Participants were not just choosing functionality; they were expressing how much decision-making they want to keep. This points to the role of \textit{user agency}, i.e., the feeling of being the one in charge, of understanding and approving what happens in the vehicle~\cite{bergstromSenseAgencyUser2022, shneiderman2021humancenteredai}. Even if full autonomy is technically possible, many users may prefer systems that preserve their ability to guide or override actions.  In this context, it remains an open question how these preferences will develop over time. If people experience clear benefits from autonomous behavior, such as saving time or increased comfort, acceptance of higher autonomy may increase. This also raises the question of how (much) to keep the human in the loop in the system design and in the user-vehicle interaction itself (e.g., via Bayesian Optimization~\cite{Jansen2025OptiCarVis}).

Our findings from the online survey further show a large variation in preferred LoA. However, because each level is predicted independently, it can also produce non-contiguous recommendations such as Levels 0, 1, and 4, which the experts found unintuitive. Predicting a contiguous range of levels as a single outcome avoids this; in our data, such a model achieved a similar ROC-AUC on unseen functions (0.61) with a lower ordinal error (mean absolute error 1.04 vs.\ 1.07 levels for CatBoost). Instead, it supports reflection and can help initiate design discussions. It may also inform later steps in user-centered design processes.

\subsection{Application in Other Areas}
While the primary focus of this work is on automotive systems, the framework we’ve developed has potential applications in various other domains. For instance, in Generative AI systems, where the level of autonomy can be adjusted depending on the complexity of the task and user preferences for control, this framework could help create more responsive and personalized systems. Similarly, in smart home automation or healthcare, autonomy levels for tasks like temperature adjustments, managing schedules, or assisting with patient care could be optimized based on FCD.

\subsection{Limitations and Future Work}
First, the classification models yielded modest performance (macro-F$_1$ 41--42\,\%, weighted F$_1$ 45\,\%), only a few percentage points above predicting every level, and close to this baseline for functions not seen during training. Although we used class weighting to address the class imbalance, performance might be improved by models that account for the ordinal structure of the LoA and predict contiguous ranges of levels, by participant-level covariates such as the propensity to trust (20\,\% of the variance in preferred LoA lies between participants), and by a larger and more diverse set of functions. Second, many hyperparameter configurations reached near-identical cross-validation scores, so the selected configuration depends on package versions. To enable better-performing models, we additionally encourage future work to make their data available. In addition, expert feedback emphasized that predictions were sometimes unintuitive, pointing to clearer explanations and more interactive visualizations. Notably, some experts raised concerns about reverse-engineering effects, where users might manipulate inputs to steer outputs, highlighting the tension between transparency and bias in predictive interfaces.

While we applied a multi-method approach, our study was limited to a specific workshop context and a single survey sample. Future work should replicate and extend the evaluation in broader use contexts and across more diverse user groups. To support this, we make our data and ML approaches openly available.

Several directions for future research emerge. First, \tool could be improved and extended based on the feedback received, for example, by adding visual explanations or allowing users to explore prediction reasoning interactively. Second, post-hoc analysis based on the demographic, personality (BFI-2-S), and Propensity to Trust questionnaires could examine how autonomy preferences differ across user groups. This could test the hypothesis that higher general trust in automation correlates with preference for more autonomous behavior and examine whether driving experience impacts the importance of function characteristics for proactivity, as existing work found differences in the perception of proactivity between more experienced and novice drivers \cite{shin2025exploratory}.

Additionally, while our work focused on identifying function-specific autonomy ranges without relying on contextual scenarios, future work should examine how contextual factors, such as driving mode, user state, or social settings, influence these preferences. This connects to the broader research on \textit{if}, \textit{how}, and \textit{when} to act proactively~\cite{nothdurft2014if, semmens2019now, wu2021learning, mathis2024approach}. Finally, the characterizing dimensions themselves may evolve over time, as vehicles become more intelligent and functions more integrated. Future studies could investigate how such dimensions shift based on vehicle capabilities, usage patterns, or emerging autonomy scenarios.


\section{Conclusion}
Our work presents a practical, data‑driven framework for matching in‑vehicle functions to appropriate autonomy levels. We distilled five autonomy levels and twelve function characteristics from expert input, surveyed 126 drivers on fourteen common vehicle functions, and trained five ML models to predict autonomy ranges from those characteristics. The resulting \tool lets designers tweak FCD ratings and immediately see autonomy recommendations, fostering transparent, user‑centered decision‑making. Expert feedback confirms its promise and highlights the need for even sharper dimension definitions and richer contextual inputs.

\section*{Open Science}
The source code and data for the ML experiments are available at \url{https://github.com/M-Colley/levels-of-autonomy-assistants}. 
All results were obtained with Python 3.13, scikit-learn 1.9.1, CatBoost 1.2.10, XGBoost 3.4.1, and TabPFN 9.2.0 using the TabPFN-v2 weights.

\bibliographystyle{ACM-Reference-Format}
\bibliography{sample-base}

\end{document}